\documentclass[11pt]{article}
\usepackage{graphicx,lscape}
\usepackage{url}
\usepackage{cite}
\usepackage{times}
\begin{document}
\pagestyle{empty}
\title{Experimental study of pedestrian flow through a bottleneck}
\author{Tobias Kretz, Anna Gr{\"u}nebohm and Michael Schreckenberg\\ 
Physik von Transport und Verkehr, Universit\"{a}t Duisburg-Essen,\\
 47048 Duisburg, Germany \\
\{kretz,gruenebohm,schreckenberg\}@traffic.uni-duisburg.de}

\maketitle

\begin{abstract}
In this work the results of a bottleneck experiment with pedestrians are presented in the form of total times, fluxes, specific fluxes, and time gaps. 
A main aim was to find the dependence of these values from the bottleneck width.
The results show a linear decline of the specific flux with increasing width as long as only one person at a time can pass, and a constant value for larger bottleneck widths.
Differences between small (one person at a time) and wide bottlenecks (two persons at a time) were also found in the distribution of time gaps.
\end{abstract}

\section{Introduction}
Bottlenecks are of interest in many systems as traffic \cite{Popkov01, Barlovic02, Kerner05, Yamamoto06}, the internet \cite{Sreenivasan06}, the ASEP \cite{Janowsky92,Klumpp04}, evolution theory \cite{Nei75} and a lot more.
A bottleneck typically denominates a limited area (in a general sense) of reduced capacity or increased demand (e.g. on-ramps on highways).
This capacity reduction can be due to a forced speed reduction (speed limit in traffic), a reduced movement probability (ASEP, tunnel effect), a reduced bias or correlation in a dynamic process (ASEP, correlated random walks), or a direct capacity reduction (networks, blocked highway lanes).
For pedestrians bottlenecks are usually formed by direct capacity reduction (door or corridor).\par
Bottlenecks are of fundamental importance in the calculation of evacuation times and other observables for buildings.
This directly implies the need to understand the phenomenons that occur in junction with bottlenecks quite well to build reliable (simulation) models of pedestrian movement.
Therefore there has been increasing interest into pedestrian flow through bottlenecks in recent years \cite{Hoogendoorn05,Kirchner03,Rupprecht05,Nagai06,Itoh02,Tajima01a,Tajima01b,Helbing03,Daamen05,Muir96,Kretz06}.
Often such research has as one aim the construction of models of pedestrian dynamics or to create data or tools to validate such models. 
One set of very basic tests that aim at giving authorities and applicants criteria to evaluate such simulation models can be found in \cite{MSC1033}.
Deciding in what situations a precise reproduction of reality is essential is one thing, knowing the reality of these situations another.
In the present situation there are much more simulation models (\cite{ASERI, BlueAdler00, ped01:Dijkstra, Egress, EVACNET, Exit89, Exodus, FAST, Helbing95, ped01:Hoogendoorn, Legion, Myriad, tgf99:Ohi, PEDFLOW, Paxport, PedGo, Nishinari04, Simulex, SimWalk, ped01:Sugiyama} to mention just a few) than empirically well investigated test scenarios.
The present study aims at closing this gap a little bit.\par
The outline of this paper is as follows.
First the scenario is described in terms of the geometrical layout.
Then the group of participants is described.
The sequence of runs - in terms of the bottleneck's widths - might also be of some importance and concludes the description of the experimental setting.
The section ``Results" begins with an analysis of the starting phase and starting effects.
From total times over fluxes, specific fluxes to the distribution of time gaps the results then proceed from macroscopic to microscopic data.

\section{Experimental Setting}
\subsection{The Geometrical Layout}
\begin{figure}[htbp]
\begin{center}
\includegraphics[width=0.8\textwidth]{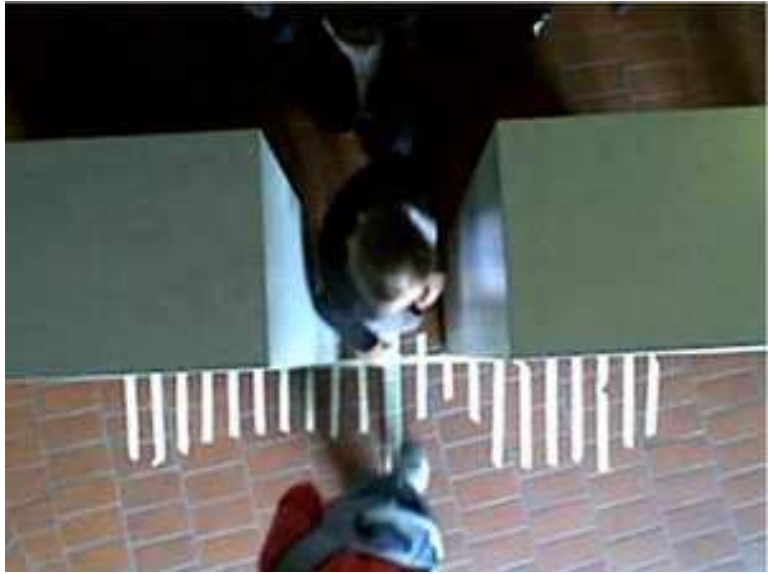}
\caption{A snapshot from the recordings at a bottleneck width of $40$ cm.}
\label{fig:floorplan}
\end{center}
\end{figure}

The experiment follows test scenario $4$ of \cite{MSC1033} but exceeds it in the amount of aspects considered.
It took place in a building at the campus Duisburg of Duisburg-Essen University from $3$ pm to $5$ pm on the $15$th of Mai $2006$. 
The bottleneck was formed by two cabinets with a height of two meters and a depth of $40$ centimeters.
Note that especially compared to the experiments reported about in \cite{Hoogendoorn05,Rupprecht05}, this bottleneck with a depth of $40$ cm is rather short, i.e. it is rather comparable to a door than a corridor.
The space in front of the bottleneck was about $4$ meter wide and $9$ meter deep with a slight increase of the width toward the back wall.
At the beginning of each run, the participants stood right in front of the bottleneck, there was no noteworthy free space between them and the bottleneck.
The process was videotaped from above (see figure \ref{fig:floorplan}) and from the side.
For analysis only the former one was used.
The time distance between two frames and therefore the time resolution is $0.033367$ seconds.
Higher time resolutions would have been possible, yet it becomes difficult at some point to distinguish between ``person has not yet passed" and ``person has passed" if the frames are too similar.
The criterion for ``person has passed" was when the first frame showed that the head of a participant fully crossed the line shown in figure \ref{fig:floorplan}.
The evaluation was done manually and not automatically by some software. In many cases it was easy for two counting people to agree about in which frame a person had passed the bottleneck. However, there also were ambiguities, where it was not easy to judge whether a participant had passed in one frame or only later in the next one.
Ten different widths of the bottleneck were examined: $40$, $50$, $60$, $70$, $80$, $90$, $100$, $120$, $140$ and $160$ centimeter.
For each reconfiguration the width was measured with a laser distance measurement device and accepted if the difference to the exact value was $2$ millimeter or below. 
The cabinets were weighed down ($>300$ $kg$) to the point that the possibility of a displacement during a run was excluded.

\subsection{The Participants}
The majority of the $94$ participants were students at Duisburg-Essen University ($32$ female, $62$ male).
Thus the group was rather homogeneous concerning age (six born before $1978$, one after $1990$, most around $1984$) and level of fitness.
Concerning body height there was a ``female frequency peak" around $170$ cm and a ``male frequency peak" around 180 cm with approximately ten guys being taller than $190$ cm.\par
The participants have been told to be vigilant and not to dawdle, but that they should not think of a competition or an emergency situation.
The homogeneity of the group must let one suspect that they are able to estimate each other's behavior quite well, which probably led to comparatively
small time gaps.
The fact that most of the participants are in the age of maximal physical fitness as well as maximal reactivity will probably have had the same effect.\par
Concerning the homogeneity of the group: one will probably quite often find groups (defined by spatial proximity) that are significantly more homogeneous than a random sample of the society.
There is most often a reason, why individuals meet and move within groups and these reasons often have a selecting effect.
In fact it is not implausible to assume that subsets of the population that are truly representative only rarely gather autonomously.
\subsection{The Sequence of Runs}
The sequence of runs is shown in table \ref{tab:script}.
\begin{table}[htbp]
\begin{center}
\begin{tabular}{c|c}
Bottleneck width&Number of runs\\ \hline
$100$ cm & 6 \\
 $90$ cm & 3 \\
 $80$ cm & 3 \\
 $70$ cm & 3 \\
 $60$ cm & 3 \\
 $50$ cm & 3 \\
 $40$ cm & 3 \\
$120$ cm & 4 \\
$140$ cm & 3 \\
$160$ cm & 3 \\
\end{tabular}
\caption{The sequence of runs.}
\label{tab:script}
\end{center}
\end{table}
For the bottleneck width of $100$ cm, six repetitions were done for two reasons: 1) The participants were to get used to the situation, especially to being filmed and 2) $100$ cm is the width mentioned in \cite{MSC1033}. For $120$ cm an extra run was done unintentionally.

\section{Results}
During the planning process the aim was to have $100$ participants.
As finally $94$ participants came to take part in the experiment, the idea was to fill the gap by telling the first six persons to walk around the cabinets in some distance and pass the bottleneck again.
However due to the noise of the crowd this did not work out well in each case and so in the following the results considering the first $80$ and - where possible - the first $100$ persons are given.\par
The most direct measurement is the total time from the first to the last participant crossing the line shown in figure \ref{fig:floorplan}.
This implies, that the total time is the sum of $n-1$ time gaps, if there are $n$ participants.
The flux then simply is the inverse of the total time multiplied by the number of persons that walked through the bottleneck in this time.
The specific flux is the flux divided by the bottleneck width.
Finally the distribution and evolution of time gaps - the time distances between subsequent persons - will be given.
But first the analysis will begin with the starting phase and therefore - to have a sufficiently detailed perspective - also with a look at the time gaps.
\subsection{The Starting Phase}\label{sub:start}
\begin{figure}[htbp]
\begin{center}
\includegraphics[width=0.8\textwidth]{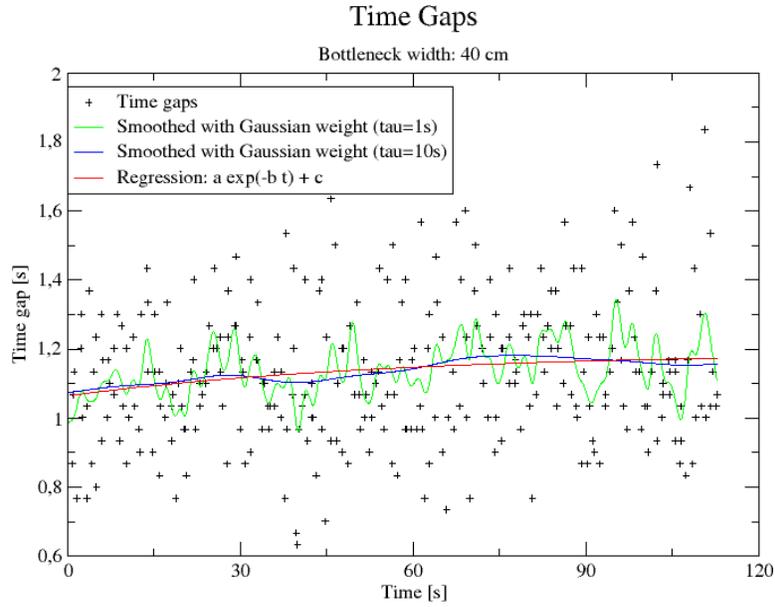}\\ \vspace{24pt}
\includegraphics[width=0.8\textwidth]{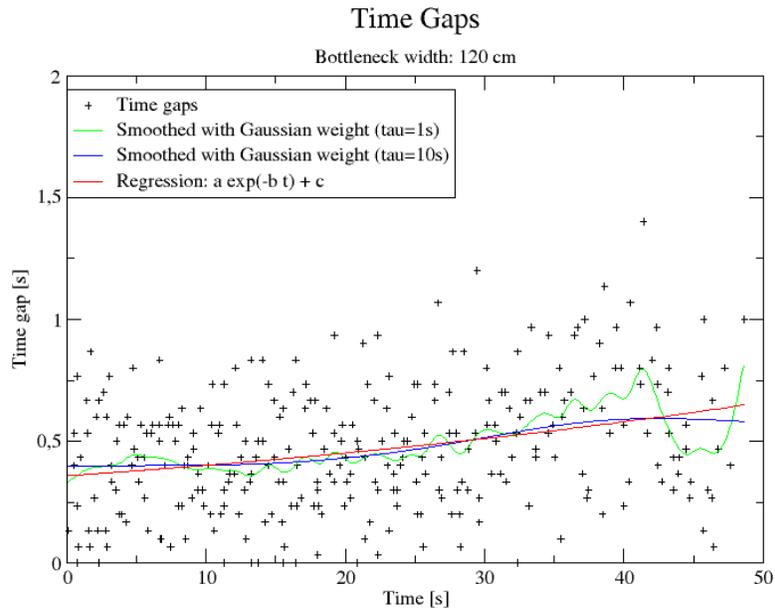}
\caption{The original time gaps data (black ``+"s) and from that data two smoothed functions (green and blue curves) and a regression following $\tilde{T}_G(t)=a \exp{(-b t)} + c$ for the widths $40$ and $120$ cm. The numerical values for $a$, $b$, and $c$ can be found in table \ref{tab:timegapsexp}. The smoothing of the blue and green curve has been done using a Gaussian weight dependent on the time distance between the considered point in time and the time of the time gap (see equation (\ref{eq:Gaussian_smooth})).}
\label{fig:time_gaps_C}
\end{center}
\end{figure}

The question is whether the process needs some time to become static or if the measured observables are the same from start to end.
A standard assumption would be an exponential relaxation of the time gaps $T_G$ following equation (\ref{eq:fit}).
\begin{equation}
\tilde{T}_G(t)=a \exp{(-b t)} + c. \label{eq:fit}
\end{equation}
The large dispersion of the data does not necessarily select the function of equation (\ref{eq:fit}) and exclude other ones.
It was chosen for reasons of simplicity as well as comparability to \cite{Rupprecht05}.
The results for the three parameters are shown in table \ref{tab:timegapsexp} and for two of them together with the original data and smoothed functions $\hat{T}_G(t)$ (following equation (\ref{eq:Gaussian_smooth})) in figures \ref{fig:time_gaps_C}.
\begin{equation} \label{eq:Gaussian_smooth}
\hat{T}_G(t)=\frac{\sum_{i=1}^{N_{max}}T_G^i\exp{\left(-\frac{(t-t(T_G^i))^2}{\tau^2}\right)}}{\sum_{i=1}^{N_{max}}\exp{\left(-\frac{(t-t(T_G^i))^2}{\tau^2}\right)}}
\end{equation}
with $t$ in steps of $0.1$ seconds, $N_{max}$ being the number of measurements and $\tau$ freely set to $\tau=1$ second or $\tau=10$ seconds.

\begin{table}[htbp]
\begin{center}
\begin{tabular}{r|rrr|rr}
Bottleneck width&$a$ &       $b$   &    $c$   & RMS	 & adjusted $R^2$\\ \hline
 $40$ cm & $-0.127$ s&  $0.018$ 1/s& $1.191$ s& $0.208$ s& $0.011$\\
 $50$ cm & $-0.086$ s&  $0.192$ 1/s& $0.986$ s& $0.180$ s&$-0.006$\\
 $70$ cm & $-0.157$ s&  $0.024$ 1/s& $0.897$ s& $0.243$ s& $0.010$\\
 $80$ cm & $-0.087$ s&  $0.099$ 1/s& $0.714$ s& $0.274$ s&$-0.005$\\
$100$ cm &  $0.006$ s& $-0.062$ 1/s& $0.511$ s& $0.296$ s& $0.021$\\
$120$ cm &  $0.255$ s& $-0.016$ 1/s& $0.103$ s& $0.246$ s& $0.086$\\
\end{tabular}
\caption{Results of a regression following equation (\ref{eq:fit}) for the temporal evolution of the time gaps.}
\label{tab:timegapsexp}
\end{center}
\end{table}

The large dispersion of measurements and the fact that no trend is recognizable concerning the dependence of $a$ and $b$ on the bottleneck width make the results for the parameters $a$ and $b$ seem to be not very reliable.
The negative value of $b$ at $100$ and $120$ cm indicate, that there is either no relaxation tendency for the first $100$ participants or that at least if there is such a tendency it is obscured by another effect.
The positive values for $a$ and the negative ones for $b$ have different causes for $100$ and $120$ cm .
While for $120$ cm this is probably caused by the tailback described at the beginning of the next section, this cannot be the cause at a width of $100$ cm, since even for the first $18$ seconds the fit revealed positive $a$ and negative $b$.
In addition, there is another local RMS-minimum for a width of $100$ cm at $a=-0.618$, $b=3.685$ and $c=0.563$ with a RMS only slightly smaller than for the result given in table \ref{tab:timegapsexp} and thus a sharp and quick relaxation on a time scale of $1/b=0.27$ s, which is less than most time gaps.
The most likely conclusion seems that for $100$ cm there is neither a relaxation tendency nor some exponential increase.
Due to the short time in which the runs for $120$ cm are completed, besides the tailback a second reason for the evolution of time gaps at $120$ cm could in principle be that the relaxation has not been completed before all participants have passed.
In this case, however, one would expect that this trend somehow can already be identified for $100$ or even $80$ cm and thus would expect different results for $80$ and $100$ cm than have actually been measured: $b$ is comparatively large for $80$ cm and $a$ is much smaller for all other widths than $120$ cm.\par
For parameter $c$ however, there is a nice trend except for the result for a width of $120$ cm.
As for positive $b$ $c$ is the static average on the long run, it is the parameter which should be easiest to fit.
$a$ and $b$ are meant to show the deviation from the static process in the beginning.
Therefore a much larger fraction of the measurements can be used to fit $c$ compared to $a$ and $b$.\par
Another possibility to estimate the time dependence of the time gaps is to watch the evolution of the average time gap between person $n-1$ and person $n$.
Averaging over six ($100$ cm) or three (rest) runs leads to far too large fluctuations to tell something about starting effects.
Therefore one has to average over a few consecutive time gaps.
\begin{figure}[htbp]
\begin{center}
\includegraphics[width=0.8\textwidth]{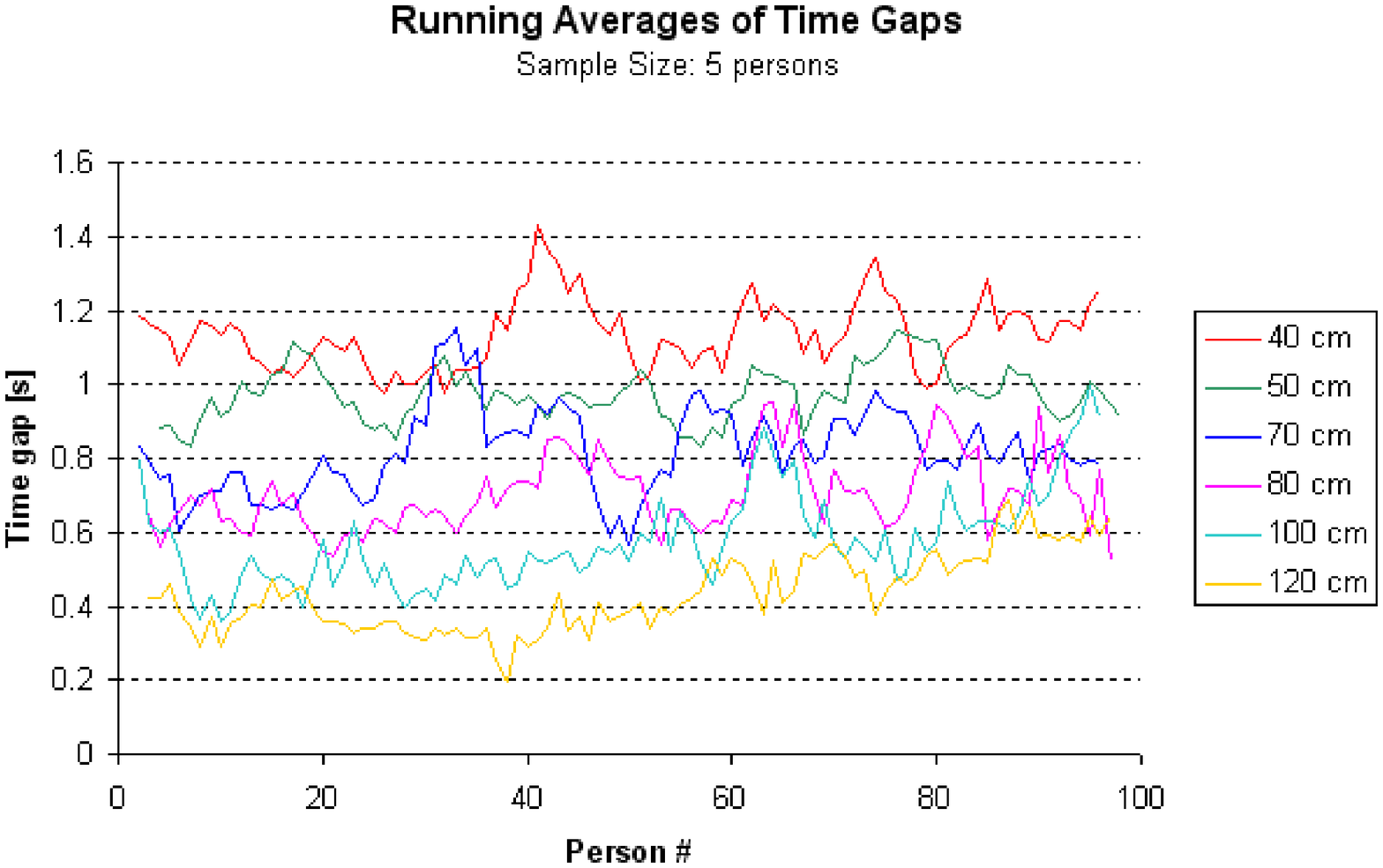} \\ \vspace{12pt}
\includegraphics[width=0.8\textwidth]{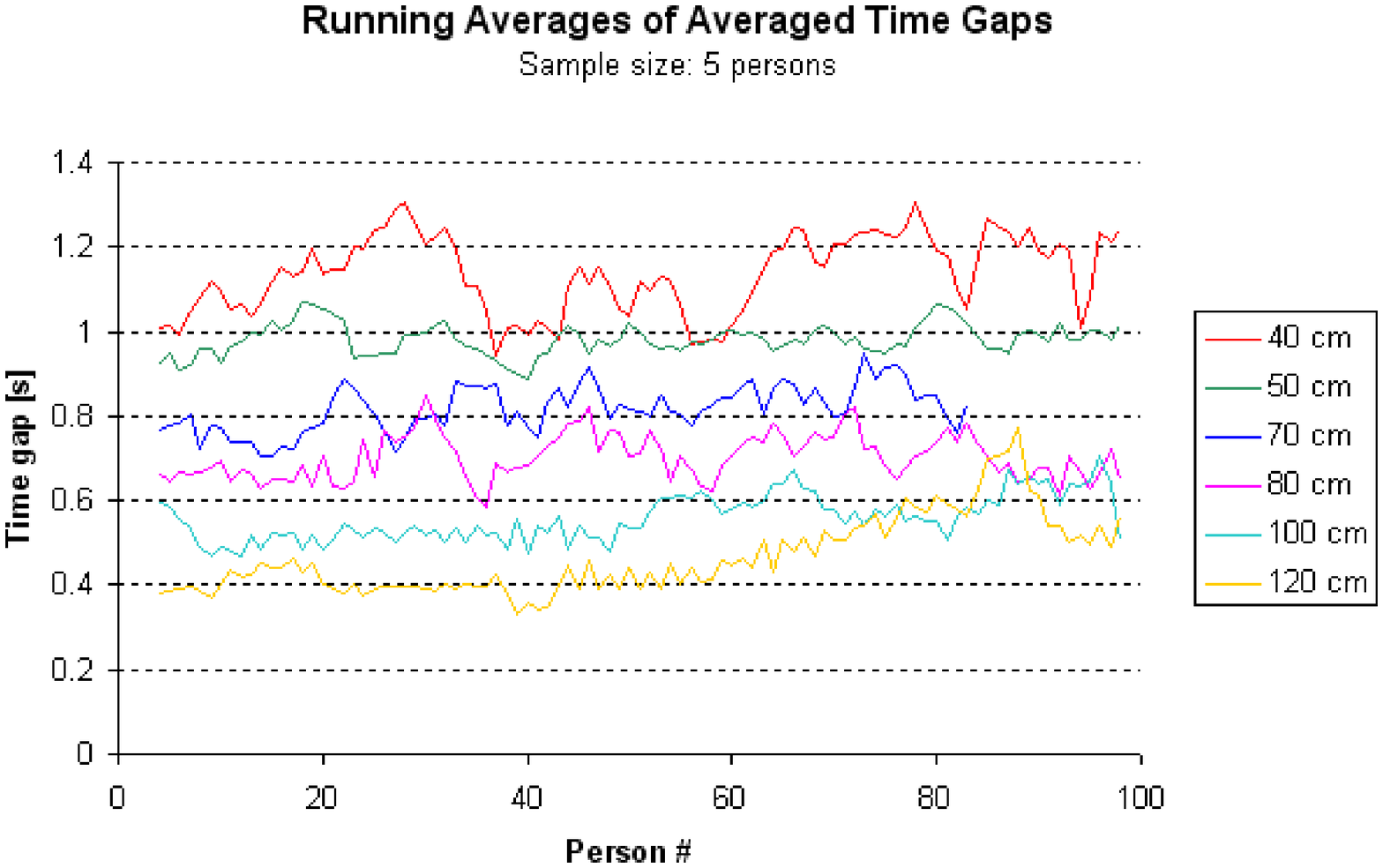}
\caption{Running averages with a sample size of five. The upper diagram includes data of just one run, while for the lower one at first an average of the $i$-th time gaps of all runs of a certain bottleneck width has been done. A data point at position ($i$/$\bar{T}_G$) therefore has the meaning, that the average of the time gaps $i-4$ to $i$ was $\bar{T}_G$.}
\label{fig:RA}
\end{center}
\end{figure}
A running average sample size of five was chosen.
For smaller sample sizes the fluctuations were too large and for larger sample sizes possible starting effects might be averaged out too much. 
Figure \ref{fig:RA} shows no starting effects with a possible exception at a bottleneck width of $100$ cm.
Or if there are trends in the beginning, they are not stronger than fluctuations that appear later on.
Since the fluctuations remain larger than possible starting effects throughout the process anyway, there seems to be no need to take starting effects into account.

\subsection{Total Times}
If - as a first assumption - one assumes a linear increase of the flux with the bottleneck width, the total time is expected to depend like 
\begin{equation} \label{eq:1} 
T\approx c/(w-w_0)+t_1
\end{equation}
on the bottleneck width $w$, with $w_0$ as minimal bottleneck width where passage is possible, $t_1$ the time for one person to pass the bottleneck, and $c$ as some constant.
If one neglects the depth of the bottleneck (a person needs no time to cross the line), one can set $t_1=0$.
This is what is assumed in this work: The total time for $N$ persons consists of $N-1$ time gaps between consecutive persons.\par
\begin{figure}[htbp]
\begin{center}
\includegraphics[width=0.8\textwidth]{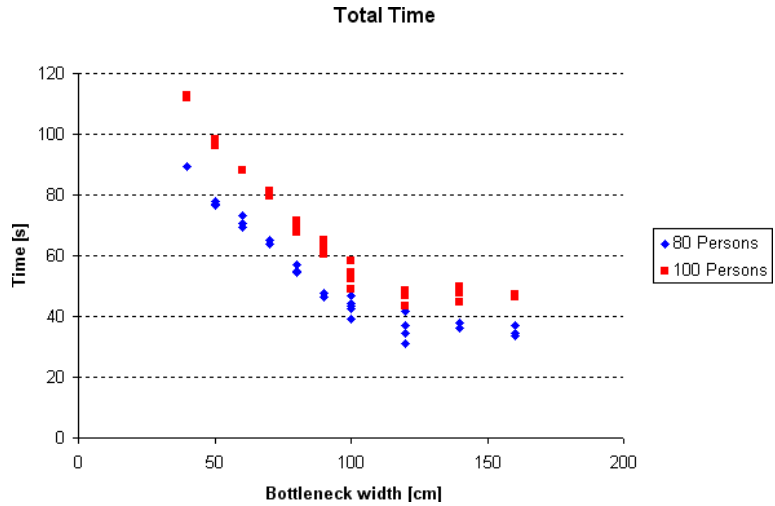}
\caption{Total times for all runs.}
\label{fig:total_times_all}
\end{center}
\end{figure}
\begin{figure}[htbp]
\begin{center}
\includegraphics[width=0.8\textwidth]{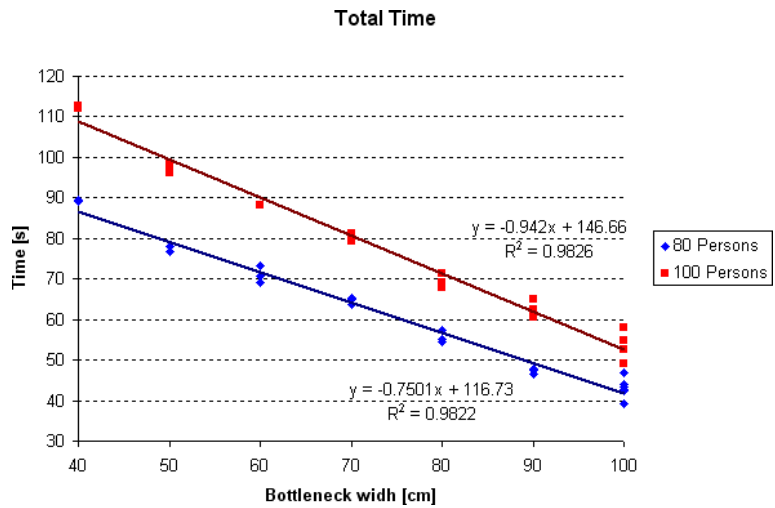}
\caption{Total times for bottleneck widths between $40$ and $100$ cm.}
\label{fig:total_times_some}
\end{center}
\end{figure}

Figure \ref{fig:total_times_all} shows a decrease of the total time until the bottleneck width reaches $120$ cm. 
That the decrease does not continue for wider bottlenecks is not only due to a normal $1/w$ behavior, which comes close to its static value, but at least partly due to participants who did not leave the area behind the bottleneck fast enough. 
The available area behind the bottleneck and the possibilities to leave this area were not sufficient to guarantee a fast efflux of the participants at those large bottleneck widths.
Therefore the flux through the bottleneck is not limited by the bottleneck itself, but by a tailback.
Figure \ref{fig:RA} very clearly shows this effect.
One can even guess quite precisely at what time the tailback reached the bottleneck.
In other words: The bottleneck stopped being a bottleneck for widths of $120$ cm and above, in a sense that the participants - due to some obstacles - couldn't leave the area behind the bottleneck fast enough to allow those passing the bottleneck to do this as fast as they could.
Figure \ref{fig:total_times_some} shows the total times for those bottleneck widths for which a linear approximation appears to be possible.
Please note that this is not a claim that there actually is a linear dependence of the total time on the bottleneck width.
This - for reasons stated above - would not be possible over the whole range of widths anyway. 
Nevertheless an almost constant decrease of the total time should not be possible in a range from $50$ or even $40$ to $100$ cm, if one assumes equation (\ref{eq:1}) to be correct, even if one assumes $w_0=20$ cm, which would be rather small.

\subsection{Fluxes}
\begin{figure}[htbp]
\begin{center}
\includegraphics[width=0.8\textwidth]{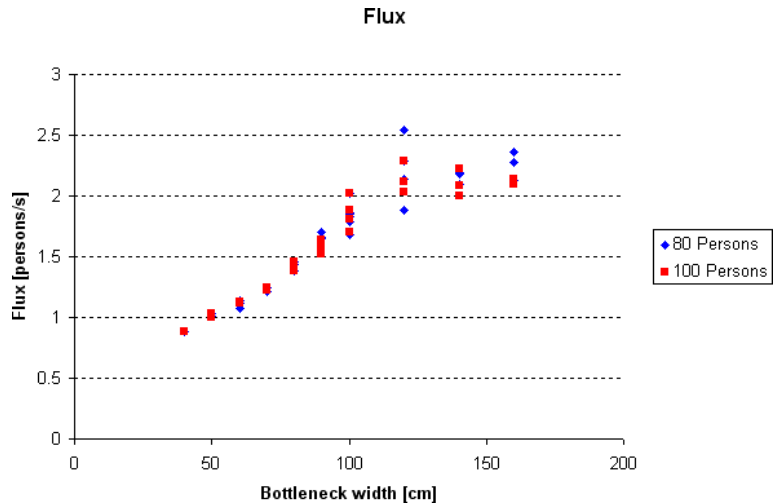}
\caption{Fluxes for all runs.}
\label{fig:flux_all}
\end{center}
\end{figure}

\begin{figure}[htbp]
\begin{center}
\includegraphics[width=0.8\textwidth]{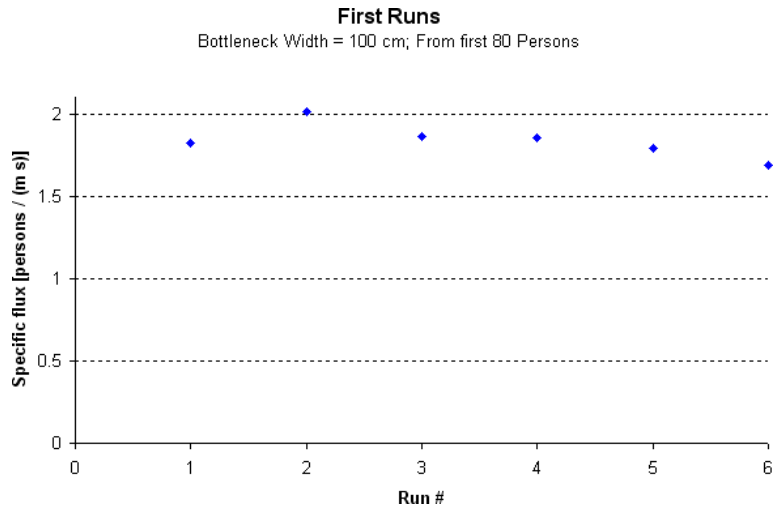}
\caption{Specific fluxes (from the ``80 persons data") for the six runs with a bottleneck width of $100$ cm in the chronological order of the runs. From this it can neither be excluded nor confirmed that there is a learning effect. So it is very hard to decide whether the wider variation of results for 100 cm compared to the other widths is due to a systematic drift during the experiment, or due to normal statistical variation and the larger number of runs.}
\label{fig:first_runs}
\end{center}
\end{figure}

Figure \ref{fig:flux_all} shows the fluxes of all runs.
One of the first things noticeable is the increase in variation for bottleneck widths of $100$ cm and above.
This has at least two reasons: 
The first is trivial, as there were six instead of three runs.
Second, one has to remember that the experiment started with a width of $100$ cm.
Maybe there has been some learning effect at the beginning of the experiment. 
The rather large variation of results for a width of $120$ cm and the development of the specific fluxes during the $100$ cm runs shown in figure \ref{fig:first_runs} neither fully support nor fully exclude this possibility.
And third there could indeed be a larger variation for widths of $100$ cm and above that is reproducible in further experiments independently of the sequence of runs or other factors.

\begin{figure}[htbp]
\begin{center}
\includegraphics[width=0.8\textwidth]{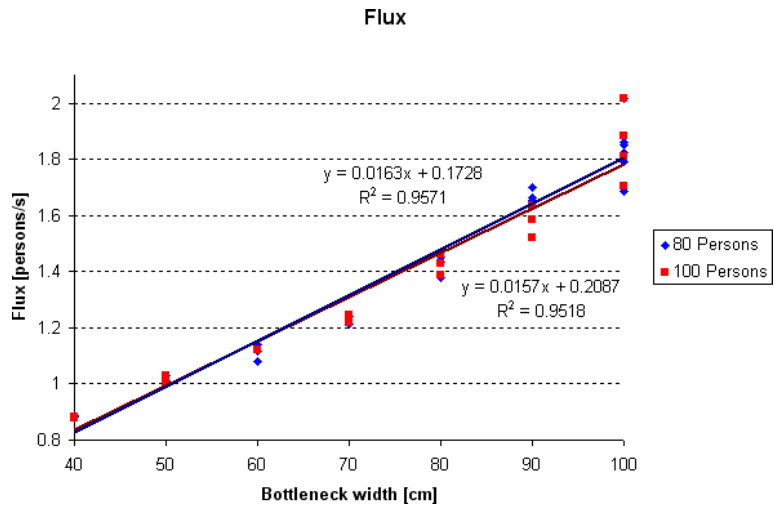}
\caption{Fluxes for bottleneck widths between $40$ and $100$ cm.}
\label{fig:flux_some}
\end{center}
\end{figure}

An interesting observation can be made in figure \ref{fig:flux_some}, where the data region is confined to bottleneck widths up to $100$ cm.
There seems to be some deviation from linearity: 
The specific flux (the slope in figure \ref{fig:flux_some}) appears to be smaller for bottleneck widths of $60$ cm and below than for larger widths. 
This will now be examined in more detail. 
Further discussion of the flux itself will be made in section \ref{sec:compare}, where it is compared to the results of other experiments.

\subsection{Specific Fluxes}
\begin{figure}[htbp]
\begin{center}
\includegraphics[width=0.8\textwidth]{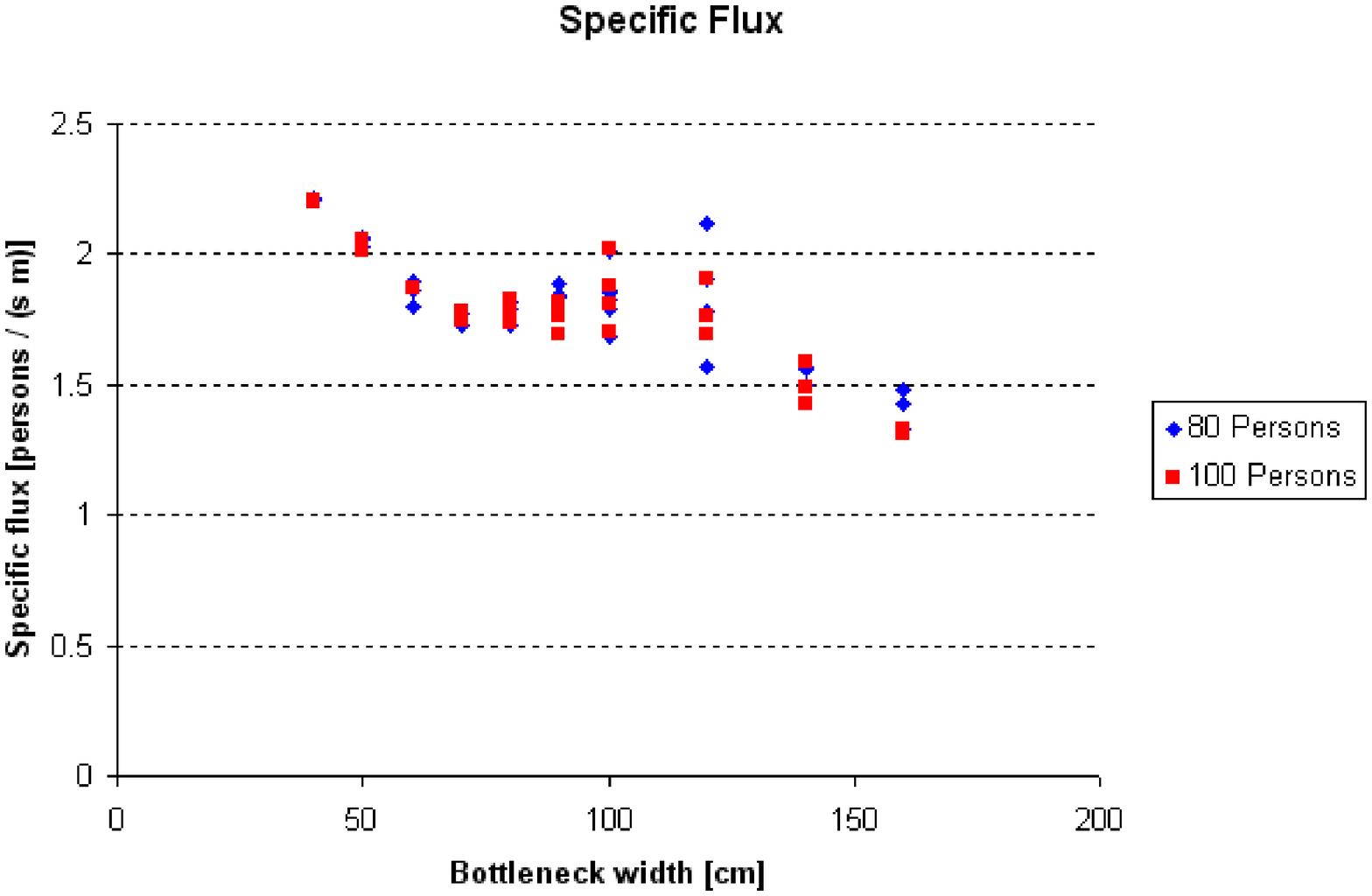}
\caption{Specific fluxes for all runs.}
\label{fig:specific_flux_all}
\end{center}
\end{figure}

\begin{figure}[htbp]
\begin{center}
\includegraphics[width=0.8\textwidth]{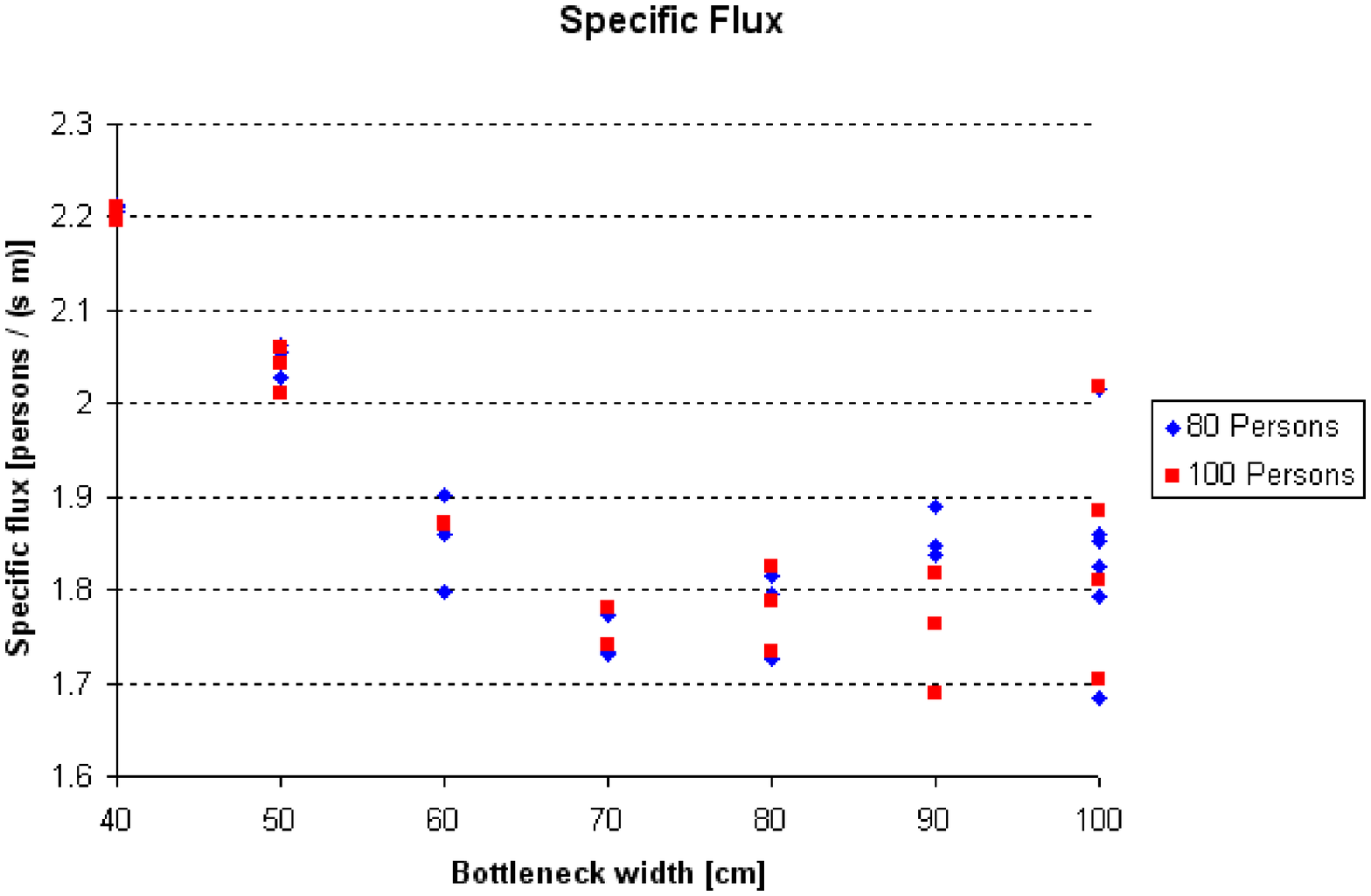}
\caption{Specific fluxes for bottleneck widths between $40$ and $100$ cm.}
\label{fig:specific_flux_some}
\end{center}
\end{figure}

\begin{figure}[htbp]
\begin{center}
\includegraphics[width=0.8\textwidth]{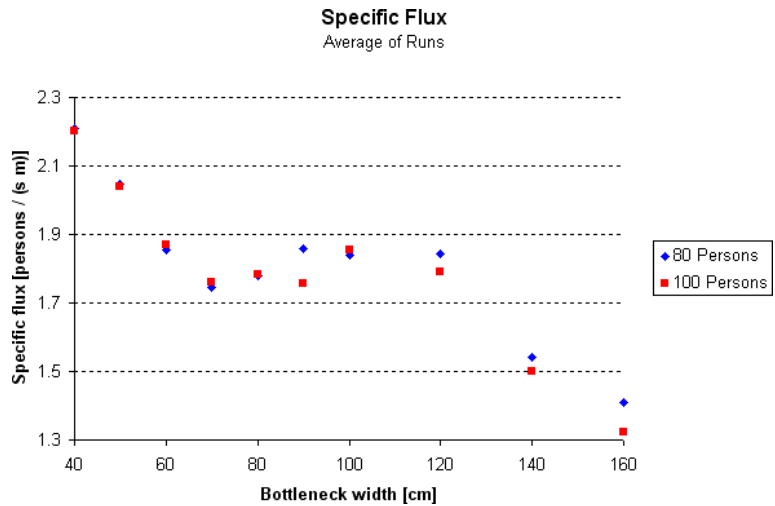}
\caption{Average of specific fluxes.}
\label{fig:specific_flux_average}
\end{center}
\end{figure}

The diagram of specific fluxes (figure \ref{fig:specific_flux_all}) shows a decrease until a width of $70$ cm.
It follows a plateau (see figure \ref{fig:specific_flux_average}) or maybe even an small increase (see figures \ref{fig:specific_flux_all} and \ref{fig:specific_flux_some}).
A non-monotonic evolution is something one would not expect.
Reasons for this could hardly be found.
Furthermore one probably would rather trust a diagram showing the averages (plateau) than one with a large dispersion of results (showing what could be interpreted as small increase).
So it seems more probable that it is a plateau with fluctuations unveiling a minimum by chance.
Why in spite of this it might be that there is an unincisive minimum is discussed in subsection \ref{sec:70}.

For small bottleneck widths of $60$ cm and below the specific flux increases since the participants actively increase their width usage efficiency by rotating their body to the side.
The reduction in the specific flux above $100$ cm on the contrary surely is caused by the tailback, but maybe also by the participants not actively taking care of efficient motion, as the task of walking through a wide ``bottleneck" may appear to be too simple to exert.

\subsection{Bottleneck Width 70 cm} \label{sec:70}
\begin{figure}[htbp]
\begin{center}
\includegraphics[width=0.8\textwidth]{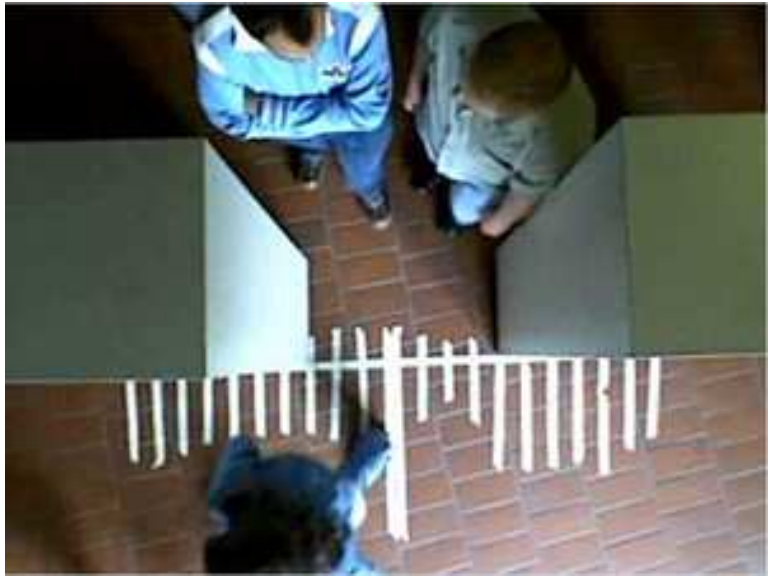}
\caption{Two participants dawdling in front of a $70$ cm bottleneck.}
\label{fig:70}
\end{center}
\end{figure}

While for all other widths it appeared that the participants did not need to communicate before passage, hesitation caused by communication appeared a number of times at a width of $70$ cm, as is shown in figure \ref{fig:70}.
This specialty of a width of $70$ cm was already noticed by several people enlisted in the organization of the experiment and later confirmed from the video footage.
The reason for this is, that at a width of $80$ cm the participants, while walking, displaced in a zipper-principle-like way:
One participant rather on the left side, then one rather on the right side and again one rather to the left.
For widths of $90$ cm or above two or more participants were able to pass at the same time, sometimes, however, this was even possible at $80$ cm.
For widths smaller than $60$ cm typically only one participant at a time passed.
The transgression from two-at-once over zipper-principle to one-by-one at a width of $70$ cm then caused those difficulties, as this appears to be the only width, where it is not obvious that typically only one person at a time can pass.
So the reason for delay is not the time needed for communicate about earlier passage.
If this was the case, the specific flux should be reduced for smaller bottlenecks even more.
The true reason is the late awareness that communication is necessary and the communication therefore falls into a time when passing would already be possible, leading to a dawdling phenomenon.\par
It is interesting to compare this phenomenon to the results of \cite{Muir96}. There the evacuation time for competitive behavior quite sharply decreses between 70 and 80 cm and the evacuation time for non-competitve behavior increases slightly between 60 and 70 cm door width. One would have to compare the recordings to tell whether these two effects can be related to the need of communication about earlier passage. But in principle one could interpret this in the following way: If there is the need to communicate about earlier passage, the participants do so in the non-competitive scenario, which leads to dawdling effects. In the competitive scenario, however, they do not. This leads to some friction-like phenomenon which compensates or even overcompensates the increased speed which follows from the higher motivation. As soon as there is no more need for communication the higher motivation in the competitive scenario can fully unfold and leads to smaller evacuation times than in the non-competitive scenario.

\subsection{Time Gaps}
\subsubsection{Distribution of Time Gaps}
The distribution of time gaps (figures \ref{fig:Dist1} and \ref{fig:Dist2}) outlines what has already been described in the last subsection.
Where the process was a strict one-by-one sequence, the distribution looks rather symmetrical with a more or less distinct maximum at the center.
The larger the width becomes, the flatter gets the distribution.
It extends more to smaller time gaps, as passing the bottleneck side by side - in principle with a time gap of zero seconds - becomes possible.
\begin{landscape}
\begin{figure}[p]
\begin{center}
\includegraphics[width=550pt]{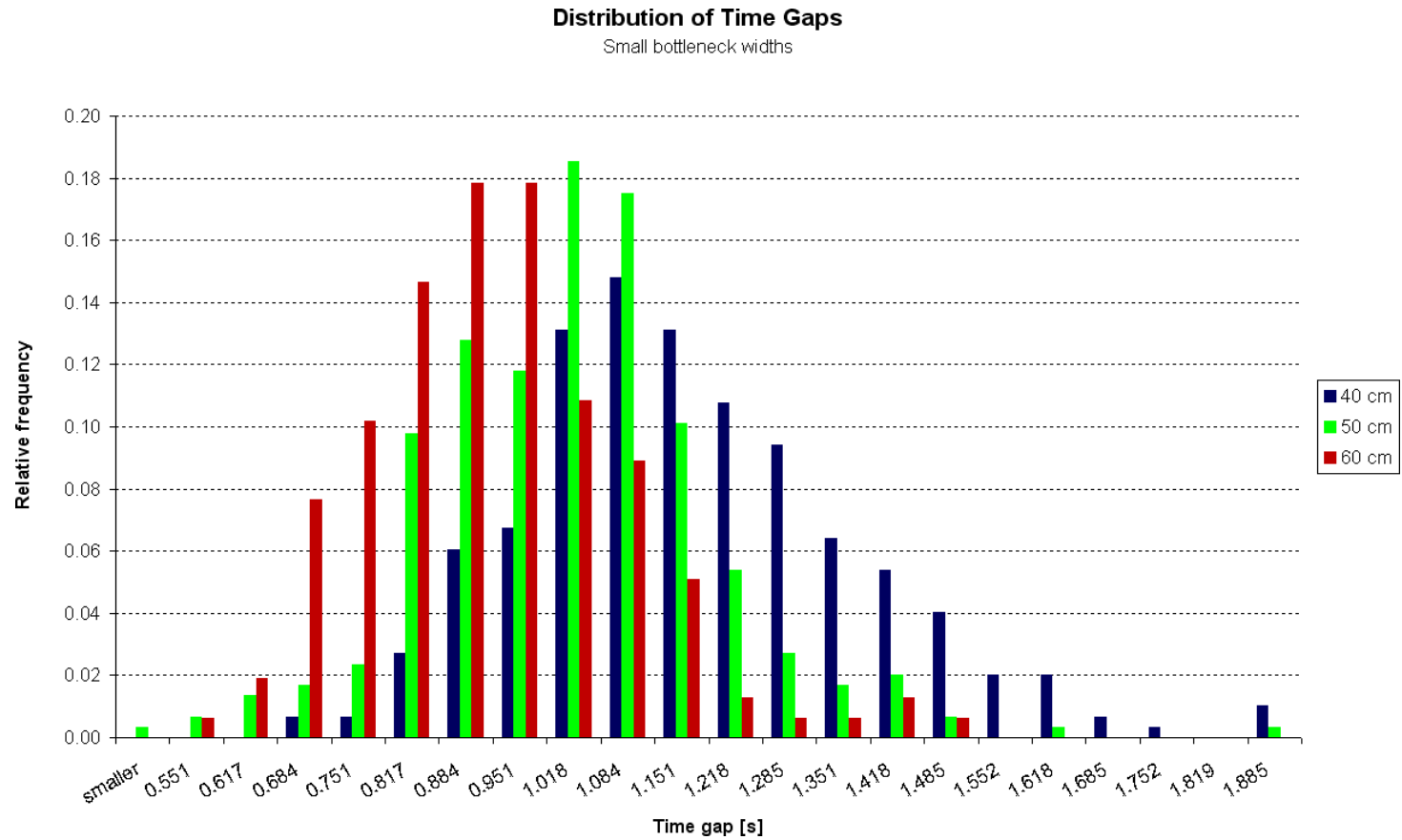}
\caption{Distribution of time gaps: small bottleneck widths. The partitioning of the time gap categories is done such that each category contains two frames.}
\label{fig:Dist1}
\end{center}
\end{figure}
\begin{figure}[p]
\begin{center}
\includegraphics[width=550pt]{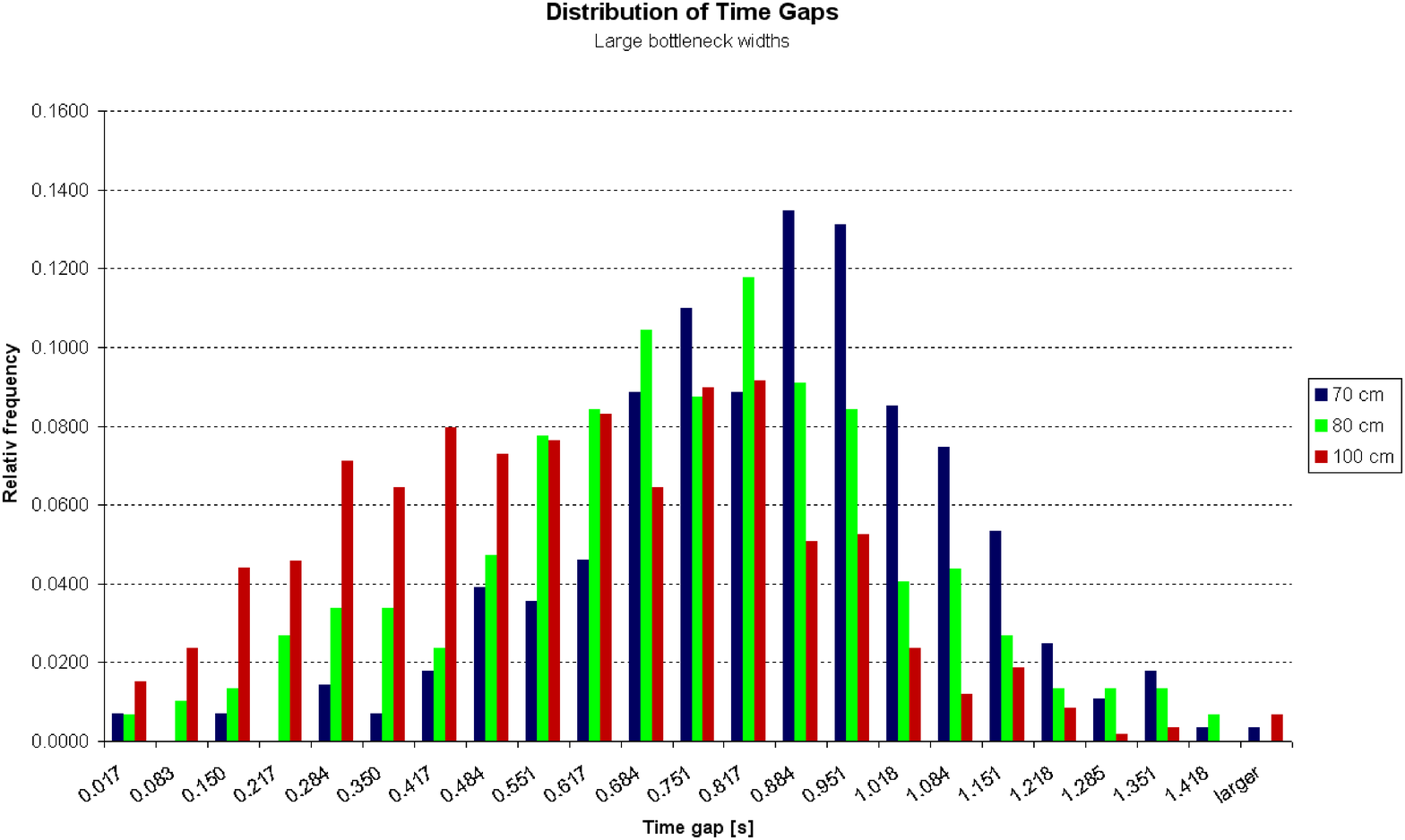}
\caption{Distribution of time gaps: large bottleneck widths. The partitioning of the time gap categories is done such that each category contains two frames.}
\label{fig:Dist2}
\end{center}
\end{figure}
\end{landscape}
Apart from this observations, the averages, standard deviations, skewnesses and kurtoses of the time gaps can be found in table \ref{tab:tg}.

\begin{table}[htbp]
\begin{center}
\begin{tabular}{c|cccc}
Bottleneck width& Average &Standard deviation&Skewness&Kurtosis \\ \hline
 $40$ cm        &$ 1.14$ s&$ 0.21$ s         & $0.59$ & $0.59$\\
 $50$ cm        &$ 0.98$ s&$ 0.18$ s         & $0.35$ & $2.63$\\
 $60$ cm        &$ 0.87$ s&$ 0.18$ s         &$-0.28$ & $4.33$\\
 $70$ cm        &$ 0.82$ s&$ 0.25$ s         & $0.11$ & $2.33$\\
 $80$ cm        &$ 0.70$ s&$ 0.27$ s         &$-0.14$ &$-0.06$\\
$100$ cm        &$ 0.56$ s&$ 0.30$ s         & $0.70$ & $2.86$\\
$120$ cm        &$ 0.47$ s&$ 0.26$ s         & $0.37$ &$-0.06$\\
\end{tabular}
\caption{Averages, standard deviations, skewnesses, and kurtoses (defined such that the kurtosis of the normal distribution equals zero) of the time gaps.}
\label{tab:tg}
\end{center}
\end{table}

The numbers of table \ref{tab:tg} exhibit...
\begin{list}{...}{\setlength{\itemsep}{0ex}}
\item as expected an inversely proportional dependence of the average time gap of the bottleneck width.
\item in parallel to the development of the specific flux a considerable variation of the product $(bottleneck width) \cdot (average time gap)$ over the different widths ($0.46$ ms for $40$ cm and $0.57$ ms for $70$ cm).
\item an agreement of the average with the regression parameter $c$ (compare table \ref{tab:timegapsexp}) that is better for small widths.
\item standard deviations that become larger as soon as the zipper-principle applies. The reason for the standard deviation at $40$ cm being larger than the ones of $50$ and $60$ cm might be, that for $40$ cm the body size plays an important role in the effort of passing the bottleneck.
\item mostly positive skewness which means - compared to the normal distribution - many small (normal behavior) and a few large time gaps (dawdling).
\item mostly positive kurtoses which means a sharper maximum than in the normal distribution. The exception at $80$ cm might be a reflection of the second peak visible in figure \ref{fig:Dist2}, indicating the two modes: two people side by side or one person walking in the middle, claiming the bottleneck alone. If one takes the values for $120$ cm for granted, the kurtoses exhibit minima each $40$ cm (at $40$, $80$ and $120$ cm). One can suspect, that a small kurtosis exhibits a ``perfect fit" of a certain number of lanes into the bottleneck. However, this needs confirmation from other experiments.
\end{list}

\section{Comparison to Similar Studies} \label{sec:compare}
There is a number of studies which are related in some point or another to this one.
These studies exceed what is reported in the following, yet the focus is on those parts that are comparable to this experiment.
In \cite{Helbing03} the results of an evacuation exercise of a classroom are reported.
The door in that experiment had a width of $50$ cm.
The maximal efflux was found to be as large as approximately $2$ persons per second.
In the present experiment at a width of $50$ cm an average flux of almost exactly $1$ person per second (compare figure \ref{fig:flux_all}) was measured and only one single time gap (compare figure \ref{fig:Dist1}) was smaller than $0.5$ seconds.
This implies that the flux observed in \cite{Helbing03} as stable flux over a few seconds appeared only as an extreme value in the present experiment.
This factor of $2$ can be explained in its tendency, yet it might still appear surprisingly large:
If the students were very young and far from being full-grown (as the focus is on the comparison with simulation results, neither the age nor the grade of the participants are stated) they would have used the available width even more efficiently than the students in the present experiment.
The second reason might be that the depth of the door might have been noticeably smaller than $40$ cm, allowing the pupils to wind around the bottleneck borders more efficiently than the students of the present study.\par
This is a suiting keyword to proceed to a comparison with \cite{Hoogendoorn05}.
In that experiment the flow through a bottleneck with a depth of $5$ meter and different widths ($1$ and $2$ meter) was measured.
From the measurements it was concluded, that the flux is a step function with respect to the bottleneck width.
This is something that could not be observed in the present experiment (compare figure \ref{fig:flux_all}).
This is probably not due to the small overlap of widths ($0.4$ to $1.2$ meter compared to $1.0$ to $2.0$ meter there) but either to the largely different depth ($0.4$ meter here compared to $5.0$ meter there) or the conclusion of a step-function is not justified.
The probability that the steps are obscured by the steps in which the bottleneck width was increased in the present experiment can be assumed to be very small.
Aside from that for a bottleneck width of $1$ meter a flux of $1.774$ persons per second is reported.
This is the result from a model calculation that is assumed to be in agreement with the measurement.
Compared to the $1.33$ persons/(m s) of \cite{MSC1033} or the $4$ persons/(m s) of \cite{Helbing03} this result is quite close to the $1.85$ (resp. $1.89$) persons per second (compare figure \ref{fig:specific_flux_average} of the present work. 
The difference can maybe be understood as stemming from the participants, who are representative for the population in the experiment of \cite{Hoogendoorn05}.\par
The study most similar to the present one is probably \cite{Rupprecht05} (participants: students, normal behavior), where between $20$ and $60$ participants walked through a bottleneck with a depth of $2.8$ meter and widths between $0.8$ and $1.2$ meter.
Aside the depth, differences to the present experiment include that the borders of the bottleneck could be overlooked by all participants and that the participants closest to the bottleneck started $3$ meter ahead of it. Thus there may have been some kind of sorting (faster ones reach the bottleneck earlier) and maybe even some kind of ``confusion" as the participants form the queue immediately in front of the bottleneck.
In any case is this difference in the initial condition most probably the reason for the difference in the distinctness of the starting phase. 
With $\lambda$ having the same meaning as the $b$ from subsection \ref{sub:start}, for the time gaps the difference with $0.16<\lambda<1.00$ compared to $0.01<|b|<0.20$ becomes evident.
Concerning the flux there is quite a good agreement between both experiments.
Especially if one considers the difference from the two ways to calculate the flux that is exhibited in \cite{Rupprecht05}.
There seems to be a tendency that for small widths the flux in the present experiment is larger, while for larger widths it is larger in \cite{Rupprecht05}.
An explanation might be, that for a short bottleneck the participants have more possibilities to increase their performance in the case of narrow bottlenecks.\par
In another study \cite{Nagai06} (participants: students, normal behavior) a significant dependence of the average flux from the initial density was found.
The maximum efflux was measured for an initial density of $5$ persons per square meter: approximately $3.3$ persons per second for a width of $120$ cm and approximately $1.7$ persons per second for a width of $40$ cm, which in the latter cases is  almost twice as large as the result of the present experiment, but - in terms of the specific flux - not as large as the flux in \cite{Helbing05} (see below for details) when a column is part of the scenario.
It is particularly interesting, that the specific flux ($4.25$ p./(m s) for $40$ cm and $2.75$ p./(m s) for $120$ cm) is larger for the smaller width, which is in agreement with the present experiment, however the relative difference is even larger in the experiment of \cite{Nagai06} than in the present one.\par
As is stated in the introduction, the participants were remembered in the beginning, that the experiment is not a competition.
Consequently, the experiment didn't exhibit any competitive characteristics.
Creating \cite{Mintz} a situation that is competitive in some respect, however, is necessary to evoke maladaptive behavior in the sense, that the density in front of the bottleneck rises to values where the flow through the bottleneck is reduced \cite{PredtetschenskiMilinski,Helbing00} significantly.
Therefore, the present experiment neither examined this regime, nor does it make any claim about the probability and requirements for a bottleneck situation going ``sub-optimal". However, contrary to the claim that panic-like situations reduce the efflux, the results of an experiment \cite{Helbing05}, where the participants were told to ``force their way through the bottleneck as fast as possible" exhibit considerably smaller average time gaps than the present experiment (compare table $2$ of \cite{Helbing05} with table \ref{tab:tg}). The smallest average time gap (door width $82$ cm) was found to be as small as $0.275$ s, which is only a fraction of $0.39$ of the overall average of $0.70$ s for a width of $80$ cm of the present experiment and it implies a specific flux as large as $4.43$ persons per meter and second. For the panic experiments $1$-$6$ (without obstacle) of \cite{Helbing05} the absolute values of the standard deviations - rather than the relative ones - of the time gaps are comparable to the one stated in table \ref{tab:tg}.\par

\section{Summary, Conclusions, and Outlook}
In this work results of a bottleneck experiment with pedestrians were presented.
Due to the participants being mostly in their twenties, comparatively large fluxes could be observed.
To be precise: all specific fluxes were found to be larger than $1.33$ persons/(m s), which is a value given in \cite{MSC1033}, comparable to those of \cite{Rupprecht05,Hoogendoorn05}, but smaller than those found in \cite{Helbing03,Helbing05,Nagai06}.
If one takes all of the results together, one finds, that the specific flux seems to increase both with the motivation as well as the initial density.
This might be in contradiction with \cite{PredtetschenskiMilinski} and the simulations of \cite{Helbing00}, where a reduction of the flux for large densities and/or over-motivation (panic) is reported.
This contradiction could be resolved by assuming, that neither the initial density in \cite{Nagai06} nor the motivation in \cite{Helbing05} were large enough to reduce the flux. 
Another possibility would be that arching and clogging exist for large initial densities and strong motivation, but they do not overcompensate the effects of initial density and motivation themselves.
This is, what is assumed in \cite{Nagai06}, where the saturation of the flux at large initial densities is attributed to clogging and not to the maximal capacity of the bottleneck.
Third, in principle there could be some maximum which has not been measured yet.
Appealing to the participants to go as fast as possible {\em as group} (i.e. minimize the total time) might help \cite{Mintz} to measure such a maximum, if it really exists.\par
Sometimes it is claimed explicitly \cite{Hoogendoorn05}, sometimes assumed implicitly \cite{MVStaettV}, that the flux is a step function of the bottleneck width.
This is something that is definitely not confirmed for bottlenecks with a smaller depth by the results of this work.
While in principle it could be that there is a step function for the flux at deeper bottlenecks and a rather linear function for bottlenecks with depths compared to the one examined here, the agreement of this work with \cite{Rupprecht05} excludes the step function.\par
The specific flux was found to be not a constant with regard to the width, but to increase with declining bottleneck width for widths smaller $70$ cm.
This can most probably be imputed to the participants moving more efficiently by rotating their body ellipse when passing the bottleneck.
This rotating of the body ellipse will probably not occur for bottlenecks where the depth is that large that more than one or two lateral steps are necessary to pass the bottleneck.\par
Only a minor if not even no starting effect could be found insofar as its amplitude is much smaller than the typical variation of measurements.\par

It would be very interesting to repeat the experiment with participants mostly older than $60$ years, with a heterogeneous group or with a significant percentage of persons carrying luggage (airports).
Another idea would be to examine the influence of the bottleneck depth, which has been $40$ cm in this work. While there have already been cited some works \cite{Daamen05,Rupprecht05} where bottlenecks with depths other than $40$ cm have been examined, the depth was never varied systematically in one single experiment.
Increasing the depth would be a transgression toward a flux in a corridor experiment, while smaller depths come closer to typical depths of doors.
And finally doing the experiment with another order of bottleneck widths - e.g. starting with $40$ cm and increasing - might lead to slightly different results.\par

Concerning model building, especially the combination of insights of the set of experiments discussed in section \ref{sec:compare} impressively demonstrates the richness of phenomenons and the number of influences that pose a challenge for any model builder.
To mention just one example: the increased efficiency at short bottlenecks demands to consider the body ellipse.
If this is implemented, a routine has to be modelled, that lets the agents in a simulation decide if they move toward a short or a deep bottleneck and if they therefore will rotate their body ellipse or not.\par

\section{Acknowledgments}
This work has been financed by the ,,Bundesministerium f{\"u}r Bildung und Forschung" (BMBF) within the PeSOS project. 
We thank Birgit Dahm-Courths, Lars Habel, Maike Kaufman, Hubert Kl{\"u}pfel, Frank K\"onigstein, Florian Mazur, and Thomas Zaksek for being supportive in the conduction of the experiment, Hubert Kl{\"u}pfel, Christian Rogsch, Andreas Schadschneider, and Armin Seyfried for some very useful discussion and of course all of the participants.

\nocite{FSS94,TGF99,PED01}
\bibliographystyle{unsrt}
\bibliography{bottleneck}

\end{document}